\documentclass[acmsmall, nonacm]{acmart}
\AtBeginDocument{%
  \providecommand\BibTeX{{%
    \normalfont B\kern-0.5em{\scshape i\kern-0.25em b}\kern-0.8em\TeX}}}

\renewcommand\footnotetextcopyrightpermission[1]{}
\setcopyright{acmcopyright}
\copyrightyear{2026}
\acmYear{2026}
\acmDOI{XXXXXXX.XXXXXXX}

\usepackage{tikz}
\usepackage{xurl}
\usepackage{float}

\usepackage{bbm}
\usepackage{pdflscape}
\usepackage{graphicx}
\usepackage{textcomp}
\usepackage{xcolor}
\usepackage{gensymb}
\usepackage{xspace}
\usepackage{booktabs}
\usepackage{tabularx}
\usepackage{multirow}
\usepackage{balance}
\usepackage{pifont}
\usepackage{caption}
\usepackage{subcaption}
\usepackage[noend]{algpseudocode}
\usepackage[linesnumbered,ruled]{algorithm2e}

\def\etal{\textit{et al.}\xspace}

\DeclareMathOperator*{\argmin}{arg\,min}

\begin{document}

\title{Beyond Boundaries: A Comprehensive Survey of Transferable Attacks on AI Systems}

% \affiliation{%

\author{Guangjing Wang}
\affiliation{%
  \institution{University of South Florida}
  \country{USA}}
\email{guangjingwang@usf.edu}

\author{Ce Zhou}
\affiliation{%
  \institution{Missouri University of Science and Technology}
  \country{USA}}
\email{cezhou@mst.edu}

\author{Yuanda Wang}
\affiliation{%
  \institution{Michigan State University}
  \country{USA}}
\email{wangy208@msu.edu}

\author{Bocheng Chen}
\affiliation{%
  \institution{University of Mississippi}
  \country{USA}}
\email{bchen5@olemiss.edu}

\author{Hanqing Guo}
\affiliation{%
  \institution{Indiana University Bloomington}
  \country{USA}}
\email{guohan@iu.edu}

\author{Qiben Yan}
\affiliation{%
  \institution{Michigan State University}
  \country{USA}}
\email{qyan@msu.edu}

\renewcommand{\shortauthors}{Guangjing Wang et al.}

\begin{abstract}
Transferable attacks allow adversarial manipulations or learned attack mechanisms constructed in one setting to compromise other AI systems, which potentially reduces attack costs and amplifies security and privacy risks. 
Yet fragmented terminology and evaluation practices hinder understanding of what transfers, why it transfers, and where transfer fails. 
This survey synthesizes representative research and develops a unified taxonomy spanning six complementary axes: cross-instance, cross-domain, cross-modality, cross-model, cross-task, and cross-deployment. 
By separating source–target configuration changes from attacker access and adaptation budgets, the taxonomy provides a common framework for comparing transfer across evasion, poisoning, backdoors, privacy inference, model stealing, and prompt-based attacks. 
We connect transfer mechanisms, including gradient alignment, shared representations, training persistence, and deployment invariances, to enhancement methods based on augmentation, disentanglement, and optimization. 
Our synthesis identifies inconsistent transfer, incompatible metrics, and gaps between simulated and physical evaluations, and establishes a research agenda for compositional transfer, configuration differences, temporal persistence, and reproducible benchmarking. 
Together, this work supports systematic investigation of shared vulnerabilities and guides transfer-aware security assessment and defense.
\end{abstract}






\maketitle

\section{Introduction}\label{intro}
Artificial intelligence (AI) systems support applications ranging from autonomous driving and speech recognition to smart manufacturing and biometric authentication, including applications of large language models (LLMs). Attacks against these systems can manipulate predictions, corrupt training, extract proprietary functionality, or disclose private information~\cite{goodfellow2014explaining,goldblum2022dataset,oliynyk2023know,hu2022membership}. Table~\ref{tab:ai-attacks} distinguishes the attack objectives between evasion, backdoor, data poisoning, model stealing, model inversion, membership inference, and prompt-based attacks.
We define transferability as a property of these attack families rather than a separate objective, and transferability describes whether an attack constructed in one setting remains effective in another~\cite{szegedy2013intriguing,gu2024survey}. 

\begin{table}[t]
\centering
\caption{Learning-based attack families considered in this survey.}
\label{tab:ai-attacks}
\vspace{-5pt}
\scalebox{0.85}{
\begin{tabularx}{1.1\linewidth}{@{}p{3.5cm}X@{}}
\toprule
Attack family & Objective \\ \midrule
Evasion & Manipulate inference inputs to induce incorrect predictions~\cite{goodfellow2014explaining,dai2018adversarial,kreuk2018fooling}. \\
Backdoor & Implant a trigger-dependent behavior through poisoned data or model modification while retaining normal behavior on clean inputs~\cite{pan2022hidden,liu2023backdoor,li2022backdoor}. \\
Data poisoning & Corrupt training data to alter learned behavior, including targeted failures or performance degradation~\cite{liu2022poisonedencoder,jiang2023color,cina2023wild}. \\
Model stealing & Replicate a model's functionality or recover its parameters~\cite{orekondy2019knockoff,liu2022stolenencoder,krishna2019thieves}. \\
Model inversion & Reconstruct sensitive attributes or training data from a model~\cite{fredrikson2015model,zhang2020secret,he2019model}. \\
Membership inference & Determine whether a sample belonged to a model's training set~\cite{shokri2017membership,hu2022membership,carlini2022membership}. \\
Prompt-based attacks & Violate instruction priority through prompt injection, or bypass safety restrictions through jailbreaking~\cite{shi2024optimization,liu2024automatic,mehrotra2024tree}. \\
\bottomrule
\end{tabularx}
}
\end{table}

\subsection{Definition of Transferable Attacks}
\label{subsec:def_trans}
In this survey, a transferable attack reuses an adversarial manipulation or learned attack mechanism constructed in a source machine learning (ML) configuration to achieve an attack objective in a target configuration. Its effectiveness depends on properties of learned models or their training processes, including learned representations, decision behavior, or information encoded through training. The attack acts through the model's input, training, parameter, or output interface. We represent a learning-based configuration as
$
Z=(\mathcal{I},\mathcal{D},\mathcal{M},\mathcal{F},\mathcal{T},\mathcal{E}),
$
where $\mathcal{I}$ denotes the input instance or instance set, $\mathcal{D}$ the data distribution, $\mathcal{M}$ the modality or input representation, $\mathcal{F}$ the learned model, $\mathcal{T}$ the task, and $\mathcal{E}$ the deployment environment. Given a source configuration $Z_s$, an attack algorithm constructs $m=\mathcal{A}(Z_s)$. 
The artifact $m$ may be a perturbation, poison, trigger, prompt, malicious model component, or learned inference mechanism. Transfer occurs when reusing $m$ achieves the attack on a target $Z_t$ that differs along at least one coordinate from $Z_s$. 
Section~\ref{bkgnd} grounds the transferable attack definition in mechanisms that can support transfer, including source--target gradient alignment, shared representations, and persistence through downstream learning.
Social engineering against humans, software exploits, and microarchitectural side channels such as Meltdown~\cite{lipp2018meltdown} are outside this definition, even when they target infrastructure hosting ML models or use ML as an attack tool.

For target configurations $\mathcal{Z}_t=\{Z_t^{(1)},\ldots,Z_t^{(n)}\}$, let $\mathcal{S}(m,Z_t)\in\{0,1\}$ indicate whether $m$ meets a specified attack-success criterion. The transfer success rate is
\[
\mathrm{TSR}(m,\mathcal{Z}_t)=\frac{1}{|\mathcal{Z}_t|}\sum_{Z_t\in\mathcal{Z}_t}\mathcal{S}(m,Z_t).
\]
The success criterion depends on the attack objective. Evasion and backdoor studies may use targeted or untargeted attack success rate (ASR)~\cite{carlini2017towards,li2022backdoor}. 
Untargeted poisoning attacks are commonly evaluated by degradation in predictive performance~\cite{biggio2012poisoning}, model-stealing attacks by fidelity (e.g., matching the predictions of the victim classifier~\cite{jagielski2020high}) to the victim model, model-inversion attacks by reconstruction quality~\cite{zhang2020secret}, and membership-inference attacks by membership advantage (e.g., the difference between the attacker's true-positive and false-positive rates~\cite{yeom2018privacy}).

\subsubsection{Threat models}
Target access and source-to-target generalization describe different aspects of an attack. Table~\ref{tab:transfer_threat_models} summarizes these distinctions. A \textit{white-box attack} can optimize directly against a known target, but establishes transfer only when its artifact is evaluated in another configuration~\cite{carlini2017towards}. A \textit{zero-query transfer attack} constructs and applies the artifact without target feedback~\cite{liu2016delving}. A \textit{query-assisted transfer attack} uses target feedback for source construction, candidate selection, or refinement~\cite{papernot2017practical,chen2023query}. A \textit{target-optimized black-box attack} repeatedly optimizes against the victim's outputs. 

\begin{table}[t]
\centering
\caption{Target access and adaptation.}
\label{tab:transfer_threat_models}
\vspace{-5pt}
\scalebox{0.85}{
\begin{tabularx}{1.1\linewidth}{@{}p{4cm}p{2.4cm}p{2.4cm}X@{}}
\toprule
Setting & Target internals & Target feedback & Transfer interpretation \\ \midrule
White-box & Available & Optional & Direct optimization; source attacks can subsequently be tested for transfer. \\
Zero-query transfer & Unavailable & None & Reuse without target adaptation. \\
Query-assisted transfer & Unavailable & Used & Reuse with reported source construction, selection, or refinement. \\
Target-optimized black-box & Unavailable & Iterative & Separate initial transfer from victim-specific optimization. \\
\bottomrule
\end{tabularx}
}
\end{table}

The transferable attack construction and execution phases may have different requirements.
For example, in an inference-time attack, such as membership inference, the attack model can be trained on shadow models in the construction phase, while still requiring outputs from the victim model to construct its input features in the execution phase~\cite{shokri2017membership}.
In a training-time attack, the adversary may instead control a subset of training data, model parameters, or pretrained components, and then observe effectiveness after victim training~\cite{goldblum2022dataset,jia2022badencoder}. 
Therefore, we note that a white-box surrogate, a poisoned-data capability, and a black-box victim interface are capabilities that can coexist within one type of attack. The threat model should be discussed in specific attack families in Table~\ref{tab:ai-attacks}.

\subsubsection{Transfer Axes}
The six axes in Table~\ref{tab:axes} identify what changes between source and target:
$
\Delta(Z_s,Z_t)=\{k\mid Z_s^{(k)}\neq Z_t^{(k)}\}.
$
For the cross-instance axis, a manipulation constructed using one set of inputs is reused on previously unseen inputs while the data distribution, modality, model, task, and deployment environment remain fixed or comparable. 
For the cross-domain axis, a new dataset indicates domain transfer when it changes the distribution, rather than merely supplying additional instances. For example, a changed class vocabulary within classification is treated as domain variation.
A cross-modality transferable attack reuses an adversarial manipulation or learned attack mechanism constructed in one sensing modality or input representation to achieve an attack objective in another. We define this axis broadly to encompass temporal or representation changes within a sensory channel, including image-to-video transfer~\cite{wei2022cross}, as well as changes between sensing modalities, such as red--green--blue (RGB) and infrared.

For the cross-model axis, independently trained weights constitute a model change even when the architecture is fixed.
For the cross-task axis, the task transfer changes the prediction objective in different tasks.
For the cross-deployment axis, a deployment change can occur with the learned model held fixed.  
We acknowledge that the axes are complementary: a classifier-to-detector attack can be both cross-model and cross-task, while a patch tested on another camera and detector can be both cross-deployment and cross-model. 

\begin{table}[t]
\centering
\caption{Six transfer axes. Multiple axes can apply to one attack.}
\label{tab:axes}
\vspace{-5pt}
\scalebox{0.85}{
\begin{tabularx}{1.1\linewidth}{@{}p{3cm}X@{}}
\toprule
Axis & Source-to-target change \\ \midrule
Cross-instance & New inputs with comparable distribution, model, and task; includes universal perturbations and triggers~\cite{huang2020universal,zhang2021survey}. \\
Cross-domain & A different data distribution within comparable modality and task~\cite{chen2019data,li2018unsupervised}. \\
Cross-modality & A different sensing modality, temporal format, or input representation~\cite{tu2022exploring,wei2022cross,zhang2024badcm}. \\
Cross-model & Different learned weights, architecture, or model adaptation~\cite{sun2022exploring,cai2022blackbox,li2020learning}. \\
Cross-task & A different prediction objective or output structure~\cite{wei2018transferable,zhang2022boosting,zeng2024cross}. \\
Cross-deployment & Different sensing hardware, preprocessing, or inference implementation~\cite{jiang2023glitchhiker,ji2021poltergeist}. \\
\bottomrule
\end{tabularx}
}
\end{table}

\subsection{Risks and Amplification Effects}
Transferability is regarded as a risk amplifier rather than a distinct class of harm. 
The importance of transferability lies in how it changes the feasibility and system-level consequences of these underlying attacks. 
High transferability can create security concerns even when the attacker lacks privileged access to the deployed target. Surrogate-crafted attacks, for example, have been shown to affect commercial application programming interfaces (APIs), content-moderation models, and perception pipelines under specific experimental threat models~\cite{cai2022blackbox,chang2023textguise,wei2023unified}. The security-relevant question is not whether transferability creates a new type of harm, but whether it allows an existing attack capability to extend beyond the system on which it was originally constructed.

The transferability property can reduce the repeated efforts, information, and computational resources required to attack additional systems.
First, it can reduce the need for repeated per-target attack construction. If one perturbation generalizes across multiple models and datasets, the marginal cost of attacking additional targets may decrease~\cite{liu2016delving,huang2020universal}. 
Second, transferability can enable attacks under more restrictive information conditions. Even when target-model parameters or implementation details are unavailable, an attacker may construct attacks on surrogate models and test whether they generalize to the target~\cite{papernot2016transferability2,suciu2018does}.
Third, transferability changes the appropriate unit of robustness evaluation. Evaluating a defense tailored to the model used during attack construction may provide an incomplete picture if related attacks transfer to other models or deployments~\cite{carlini2017towards,tramer2017ensemble}. 
Therefore, studying transferability motivates cross-instance, cross-domain, cross-model, cross-task, cross-modality, and cross-deployment evaluation.

\begin{figure}
    \centering
    \includegraphics[width=0.7\linewidth]{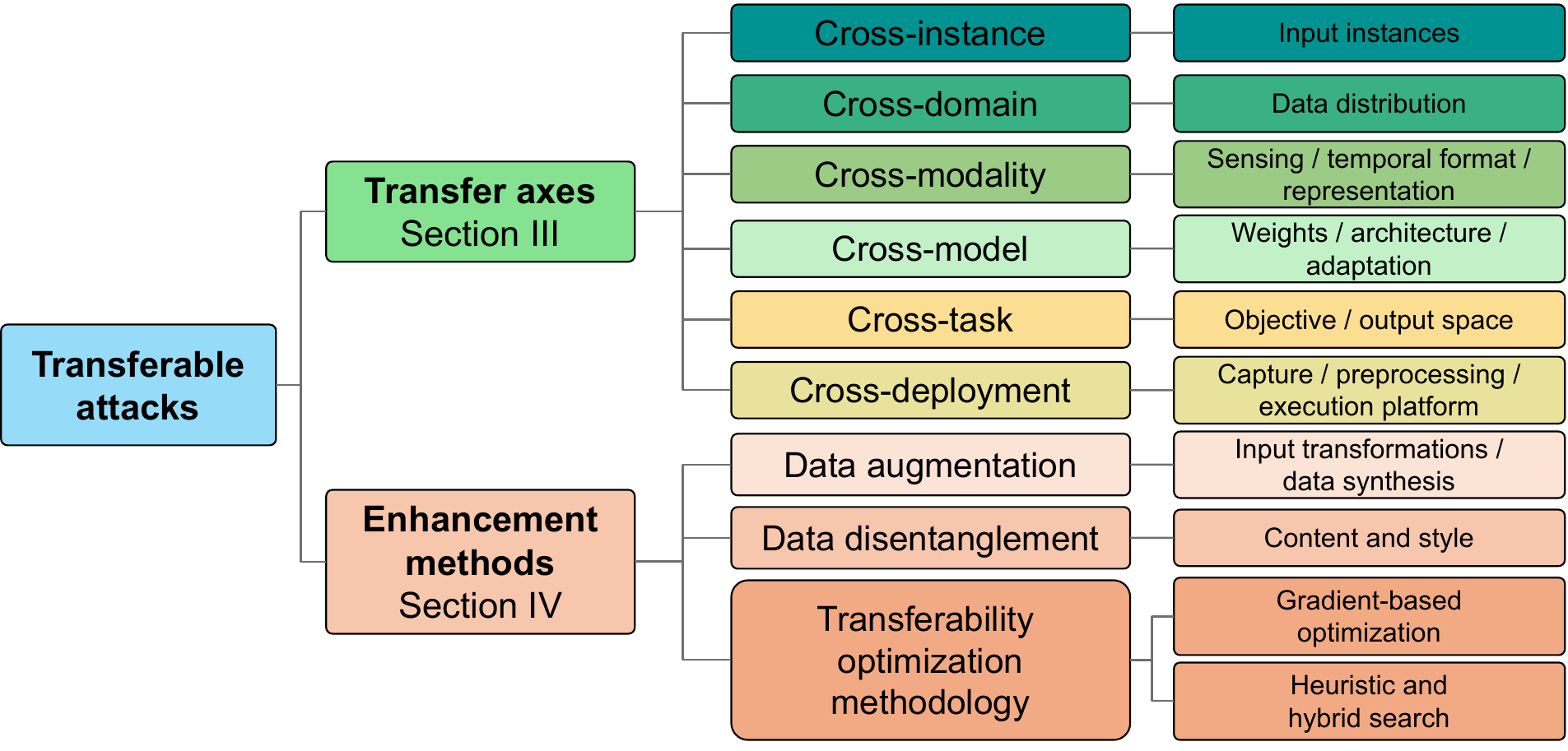}
    \caption{Transferable attacks across six transfer axes and enhancement categories.}
    \label{fig:map1}
    \Description{An overview diagram branches into six transfer axes: cross-instance, cross-domain, cross-modality, cross-model, cross-task, and cross-deployment. A second branch organizes enhancement methods into data augmentation, data disentanglement, gradient-based optimization, and heuristic search.}
    \vspace{-15pt}
\end{figure}

\subsection{Literature Selection and Scope}
We conducted a structured literature search covering publications available through August 2026. 
Searches were performed in August 2023 and August 2026 in IEEE Xplore, ACM Digital Library, SpringerLink, ScienceDirect, and relevant security, machine learning, vision, language, and speech proceedings. Search terms combined "transferable attack", "attack transferability", "surrogate model", "universal adversarial", the six transfer axes, and attack-specific terms such as "transferable backdoor" and "transferable jailbreak". 

The initial pool contained about 350 candidate publications. 
After removing 78 duplicate records, 272 papers were screened based on titles and abstracts.
For the main survey synthesis, we target a learned model or pipeline and evaluate reuse beyond its source configuration, or explain the conditions under which such reuse succeeds or fails. We prioritize representative mechanisms, empirical transfer evaluations, and foundational insights. 
Published versions are preferred where available.
Conventional cache, timing, power, or cryptographic side-channel attacks that recover secrets through implementation leakage are excluded. 
Physical sensor manipulations are included when their evaluated objective is to compromise a downstream learned pipeline.

\begin{table}[t]
\centering
\caption{Representative surveys grouped by attack family and application scope.}
\label{tab:survey}
\vspace{-5pt}
\scalebox{0.85}{
\begin{tabularx}{1.1\linewidth}{@{}p{4.5cm}p{5cm}X@{}}
\toprule
Focus & Typical scope & Representative surveys \\ \midrule
Evasion & Vision, networks, graphs, language & \cite{badjie2024adversarial,wang2023adversarial,sun2022adversarial,zhang2020adversarial,rosenberg2021adversarial,akhtar2018threat} \\
Backdoors & General learning and language & \cite{cheng2025backdoor,zhang2024backdoor,li2022backdoor,li2023backdoor,goldblum2022dataset,kaviani2021defense} \\
Data poisoning & Training and recommendation & \cite{chaalan2024path,nguyen2024manipulating,cina2023wild,tian2022comprehensive,wang2022threats} \\
Membership inference & Privacy and federated learning & \cite{bai2024membership,hu2022membership,hu2023defenses,zhang2021privacy} \\
Model inversion & Reconstruction and privacy & \cite{xu2024large,zhang2022survey,liu2021machine,dibbo2023sok} \\
Model stealing & General models and graphs & \cite{zhang2024survey,oliynyk2023know,gencc2023taxonomic,gong2020model} \\
Universal / transferable evasion & Instance or model transfer & \cite{zhang2021survey,gu2024survey} \\
\textbf{Transferable attacks} & Multiple families and modalities & \textbf{This survey}\\
\bottomrule
\end{tabularx}
}
\end{table}

\subsection{Contribution and Organization}
Existing surveys, as shown in Table~\ref{tab:survey}, organize attacks by family or application, including evasion, backdoors, poisoning, privacy inference, and stealing. Surveys of universal perturbations~\cite{zhang2021survey} and transferable adversarial examples~\cite{gu2024survey} provide deeper coverage of particular transfer settings. We connect these perspectives through six axes that apply across attack objectives. The survey contributes a consistent taxonomy for classifying transfer, a synthesis of mechanisms and enhancement methods, and an assessment of evaluation limits and open problems.

\figurename~\ref{fig:map1} outlines the survey, connecting the six transfer axes with the methods used to improve attack generalization. Section~\ref{bkgnd} reviews the discovery of transferable attacks, factors associated with transferability, and threat actor assumptions and limitations. 
Section~\ref{dimension} organizes transferable attacks across instance, domain, modality, model, task, and deployment changes. Representative studies illustrate what transfers in each setting, how transfer is evaluated, and where multiple axes overlap. Section~\ref{method} analyzes enhancement methods based on data augmentation, disentanglement, gradient-based optimization, and heuristic or hybrid search. Finally, Section~\ref{future} identifies research priorities for reproducible evaluation, comparable benchmarks, and defenses that remain effective across models and deployment conditions.

\section{Related Work on Transferability}\label{bkgnd}
\subsection{Discovery and Development of Transferable Attacks}
Transferability emerged when adversarial examples crafted for one neural network also fooled networks with different architectures or training data~\cite{szegedy2013intriguing}. Goodfellow~\etal\cite{goodfellow2014explaining} linked this behavior to locally linear responses and introduced the fast gradient sign method (FGSM). Papernot~\etal\cite{papernot2016transferability2} showed how adversary-trained substitutes enable attacks on unseen targets. Liu~\etal\cite{liu2016delving} studied targeted transfer using model ensembles. Carlini and Wagner~\cite{carlini2017towards} subsequently showed that high-confidence transferable examples could defeat defensive distillation.

Later methods sought to reduce dependence on a single surrogate. Momentum~\cite{dong2018boosting}, translated or resized inputs~\cite{dong2019evading,zou2020improving}, and model augmentation~\cite{li2020learning} improve transfer in studied settings. Research also examined attacker knowledge~\cite{suciu2018does}, local loss smoothness~\cite{wu2020towards}, and representation similarity~\cite{liang2021uncovering}. Recently, the phenomenon extends beyond classifier evasion. Vision--language model attacks exploit aligned features~\cite{wang2024transferable}, while NOTABLE~\cite{Mei2023NOTABLETB} transfers backdoor behavior across downstream natural language processing (NLP) tasks and prompting strategies. These extensions motivate examining both the transferred artifact and the configuration changes.

\subsection{Factors Associated with Transferability}
\subsubsection{Gradient geometry}
Gradient geometry helps explain cross-model evasion. Demontis~\etal\cite{demontis2019adversarial} identify the \emph{target's input-gradient magnitude, source--target gradient alignment, and surrogate-loss variability} as relevant factors.
Gradient magnitude affects attack transferability, but magnitude alone cannot determine whether the transferred attack achieves either attack efficacy or victim targeting.
Aligned gradients can provide directions that increase loss on different models to achieve cross-model attack. A local approximation clarifies the role of gradient alignment. Let $\ell_s$ and $\ell_t$ denote source and target losses for the same prediction objective. For a small perturbation $\delta$, the first-order target loss change is approximately $\nabla_x\ell_t(x)^\top\delta$. An untargeted source update aligned with $\nabla_x\ell_s(x)$ is more likely to raise target loss when the two gradients point in similar directions~\cite{demontis2019adversarial}.

Transfer-enhancement methods act on different parts of gradient alignment. A surrogate ensemble seeks directions useful for several source models~\cite{liu2016delving,dong2018boosting}. 
Input transformations seek directions that remain effective when the input is resized, translated, or otherwise varied~\cite{xie2019improving,dong2019evading}. 
Feature objectives move optimization away from a particular output head and toward representations reused across architectures or tasks~\cite{inkawhich2019feature,wang2021feature}. 
Neighborhood stability objectives reduce dependence on narrow, source-specific optima by favoring locally stable gradient directions across nearby inputs or models. In this way, achieving neighborhood stability suppresses model-specific components that are less likely to align with the victim model~\cite{wu2020towards,qin2022boosting,Long_2025_ICCV}.
However, we note that neighborhood stability does not mathematically guarantee source–target gradient alignment.
In addition, unstable surrogate behavior can reduce transferability.

\subsubsection{Shared representation}
Shared representations and training procedures provide additional pathways to achieve transferable attacks.
Similar inductive biases can align learned features~\cite{conwell2024large}, and shared backbones can influence transfer more than task-specific detection modules~\cite{li2025importance}. Ensembles, feature objectives, augmentation, and neighborhood optimization attempt to exploit the shared representation while limiting overfitting to a source model. 

By contrast, training-time attacks depend on a different form of shared representation. A backdoor needs to retain its trigger--output association after the victim trains or adapts the model~\cite{wang2020backdoor,jia2022badencoder,Mei2023NOTABLETB}. 
A successful poison affects the learned behavior after the altered data are consumed~\cite{biggio2012poisoning,liu2022poisonedencoder}. 
Shared pretraining, feature reuse, and compatible training objectives offer plausible connections to such transferable attacks~\cite{jia2022badencoder,liu2022poisonedencoder}.
Yet, trigger persistence through adaptation is a separate requirement from transferability across input instances~\cite{wang2020backdoor,li2022backdoor}.

\subsubsection{Information geometry}
Information geometry offers another account of transferability in jailbreaking attacks.
Instead of asking "How far did I move in the model's parameter coordinates?", information geometry asks "How much did that move change the model's probability distribution over outputs?"
RIGJ~\cite{wang-etal-2026-enhancing-transferability} assumes that different LLMs are trained to model similar natural-language dependencies, so their output-distribution spaces contain some shared statistical structure.
With this assumption, optimizing relative to output-distribution geometry may capture more transferable directions than optimizing according to model-specific embedding coordinates.
However, RIGJ does not prove that independently trained LLMs share equivalent information geometry.

We note that shared architectures or modalities do not guarantee transfer. Meade~\etal\cite{meade-etal-2025-investigating} find poor, inconsistent adversarial-trigger transfer across 13 LLMs, with outcomes affected by alignment strategy. 
SoundBreak~\cite{hussain-etal-2026-soundbreak} similarly reports that perturbations causing severe source-model degradation transfer poorly across trimodal models and audio encoders, which suggests that effective adversarial directions may remain model- or encoder-specific.

\subsection{Mitigation Across Attack Families}
We examine selected defenses against shared inference-time vulnerabilities, malicious behavior that persists through training, and information exposed by model outputs. 
For evasion, ensemble adversarial training incorporates perturbations from external models to improve resistance to vulnerabilities shared with surrogates. 
Meanwhile, ensemble adversarial training requires control of training and access to surrogate models and increases training cost~\cite{tramer2017ensemble}. 
Input reconstruction and detection methods instead modify or reject transferred inputs through the deployment pipeline. 
They require representative clean data for training or calibration and can introduce latency, input distortion, and false rejection~\cite{meng2017magnet}. 
For poisoning and backdoors, data filtering and model repair target contaminated samples and persistent malicious associations. 
Data filtering requires access to training data, whereas pruning and fine-tuning generally require model weights and trusted clean examples. These interventions may degrade benign performance and should be evaluated against adaptive attackers who redesign poisons or triggers to circumvent the defense~\cite{goldblum2022dataset}. 

For model stealing, output restrictions and query controls limit feedback available for constructing substitutes, which mitigates the stealing attacks. 
However, label-only extraction demonstrates that withholding confidence scores alone is insufficient~\cite{oliynyk2023know}. 
For membership inference attacks, differentially private training limits individual training records' influence on the released model, helping mitigate membership inference and record-level leakage in model inversion. However, its use requires control over training, an explicit privacy budget, and evaluation of the privacy--utility tradeoff~\cite{hu2022membership}.
For prompt-based attacks, input screening filters malicious prompts, whereas alignment training aims to reduce unsafe responses. These defenses require control over the input-processing pipeline and model training, respectively, and should be evaluated using transferred prompts while reporting refusal rates on benign requests~\cite{meade-etal-2025-investigating}.
Across attack families, defense evaluations should distinguish unchanged attacks transferred to held-out targets from adaptive attacks tailored to the full defense pipeline. The evaluation should report attack-construction and target-query budgets, and assess performance on benign inputs.

\subsection{Threat Actor Assumptions and Failure Modes}

\subsubsection{Surrogate--target similarity}
Attackers often assume that public data, architectures, or preprocessing approximate the target~\cite{papernot2016transferability2,suciu2018does}.
The assumption is reasonable because, in practice, vendors converge on shared datasets (e.g., ImageNet-style pretraining; Common Crawl for LLMs), canonical augmentations, and widely adopted toolchains. This creates aligned representations and similar gradient geometry. Attackers can replicate target model outputs sufficiently to enable successful attack transfer by training surrogate models on publicly accessible datasets.
Shared training data or pretrained features can support surrogate--target similarity~\cite{liang2021uncovering,mao2022transfer}. In studied black-box settings, attackers can train surrogates on public data and reuse their adversarial examples against unseen models~\cite{liu2016delving}.
Model extraction can further improve this approximation through API feedback~\cite{correia2021copycat,kariyappa2021maze}.
However, proprietary components, distribution shifts, and differences in preprocessing can undermine surrogate--target similarity~\cite{mao2022transfer}. In black-box deployments, undisclosed architectures and training configurations, including optimizer choices and hyperparameters, may further limit how closely a surrogate approximates the target~\cite{suciu2018does,mao2022transfer}.

\subsubsection{Inductive-bias overlap}
Related model families may learn similar features, particularly when they share tasks, objectives, or pretraining~\cite{conwell2024large}.
The assumption is reasonable in that production models cluster around a small set of families, and real systems vary across sensors but preserve task semantics~\cite{inkawhich2019feature,liang2021uncovering}. This shrinks the hypothesis space and increases gradient or feature alignment. For instance, different modalities such as image and LiDAR both target the same event, which share similar inductive biases in latent space.
However, differences in architecture and training data can yield modality-specific vulnerabilities that limit transfer~\cite{suciu2018does,mao2022transfer}. 
For example, camera and LiDAR models observing the same object need not share aligned representations or adversarial directions. 
Optimization over transformations can improve robustness to selected input variations~\cite{athalye2018synthesizing}, but does not by itself establish cross-modality transfer, especially when target sensing conditions or data distributions differ from those assumed during attack construction.

\subsubsection{Shared training and physical stability}
Many studies assume that adversaries and victims share similar training dynamics, which describes how a model's parameters, representations, and behavior evolve during training.
For instance, attackers can share training dynamics, including data, model checkpoints, or objectives that cause the same correlations.
The assumption is reasonable because reproducible ML favors shared recipes: loss functions, augmentations, epochs, and even public data mixtures and pre-trained foundation models.
Training-time transfer studies often exploit shared data, pretrained checkpoints, or learning objectives~\cite{wang2020backdoor,jia2022badencoder,liu2022poisonedencoder}. 
Reusing these components can preserve attacker-induced features or associations in downstream models~\cite{jia2022badencoder,liu2022poisonedencoder,Mei2023NOTABLETB}. 
However, in practice, fine-tuning and changes to training conditions can weaken the malicious association in shared training dynamics~\cite{li2022backdoor,goldblum2022dataset}. More importantly, the victim's data distribution is often inaccessible~\cite{alecci2023your}, which limits how closely an adversary can reproduce its training conditions.

Physical attacks additionally assume that the physical channel preserves key adversarial cues across devices and viewpoints within practical ranges.
This assumption is reasonable given that consumer optics, audio processing chains, and image signal processing (ISP) stacks are often standardized.
Physical studies test this assumption across capture conditions such as viewpoint, distance, and illumination~\cite{eykholt2018robust,zhao2019seeing}. 
However, resizing and audio compression can alter the received manipulation~\cite{xiao2019seeing,zhang2025pta}, while sensor-path attacks can perturb camera inputs~\cite{ji2021poltergeist}. 
As a result, studies of cross-deployment attacks generally need to evaluate the attack across specific perception pipelines and physical configurations~\cite{cao2021invisible,tu2022exploring}. Such evaluations help determine whether a manipulation remains effective after undergoing changes in real-world conditions.

\section{Transferable Attack Taxonomy}\label{dimension}
\begin{figure}
    \centering
    \includegraphics[width=0.8\linewidth]{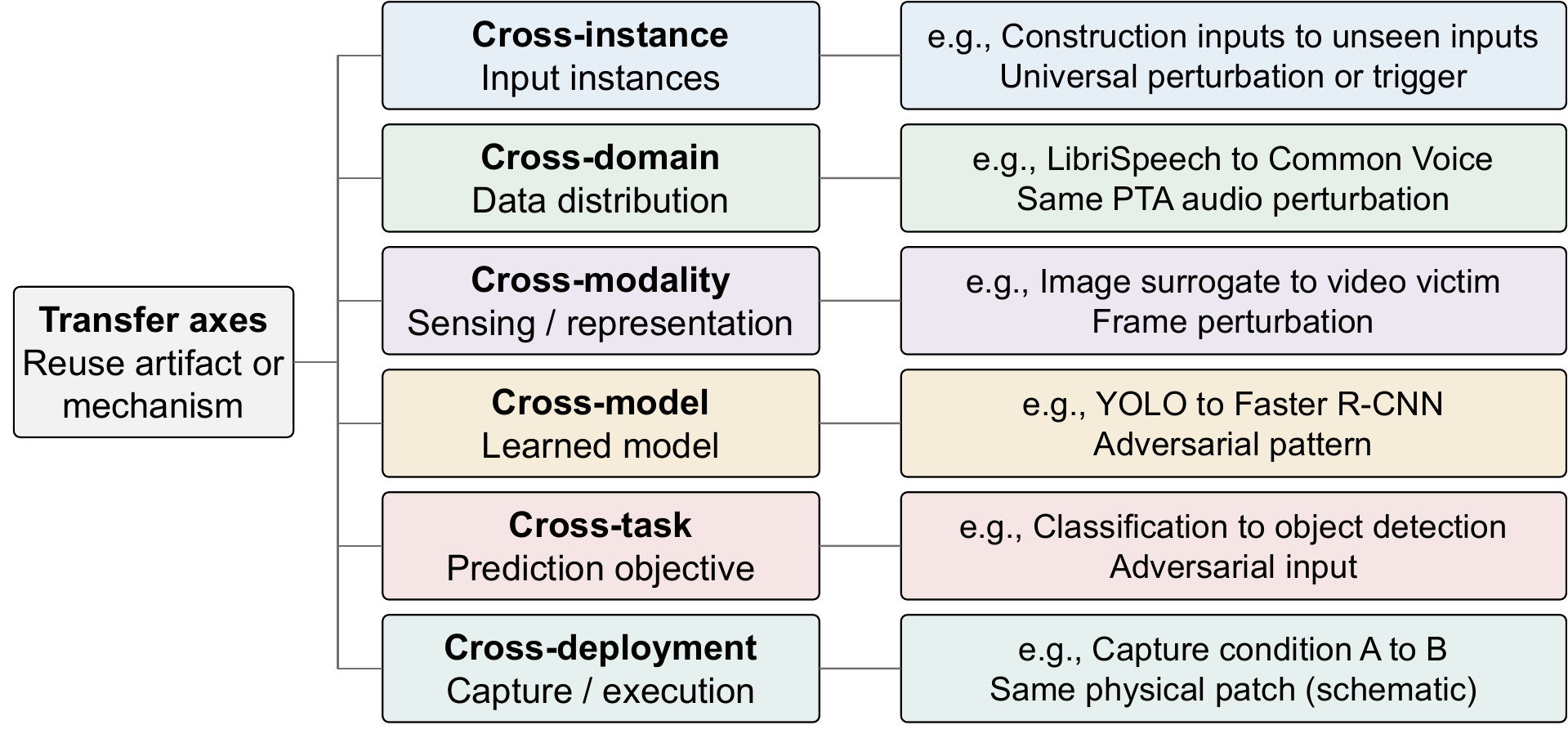}
    \caption{Transfer axes and examples in each axis.}
    \label{fig:axes}
    % \vspace{-10pt}
\end{figure}

As shown in \figurename~\ref{fig:axes}, we identify transfer axes and corresponding examples. We note that a transferable attack study can support multiple axes, and applying a methodology separately to several settings does not establish that its artifact is transferable.

\subsection{Cross-Instance Transfer}
Cross-instance transfer reuses a manipulation or transformation on new inputs under a comparable model, domain, and task.
In cross-instance attacks, construction data are used to develop the attack, whereas application inputs are previously unseen instances on which the attack is reused.
An attack may be optimized over many source examples to produce a single reusable artifact. 
Conversely, an artifact optimized on one example may be selected from several candidates using additional examples.
In particular, universal perturbations are optimized on one or more source inputs rather than on a single instance~\cite{moosavi2017universal,huang2020universal,zhang2021survey,Wallace2019UniversalAT}.
Accordingly, evaluating cross-instance transfer requires distinguishing inputs used for attack optimization and candidate selection from held-out inputs used for testing. For class-specific perturbations, evaluations should report attack effectiveness on the targeted class and effects on other classes separately. These reporting principles apply across attack forms, including image patches, audio patches, and token triggers.

Cross-instance transferability is widely explored in computer vision.
For instance, Moosavi-Dezfooli~\etal\cite{moosavi2017universal} accumulate perturbations over sampled images. 
By contrast, data-free methods exploit intermediate features~\cite{mopuri2017fast} or generated class representations~\cite{mopuri2018ask}.
Class-specific universal perturbations induce misclassification of inputs from a selected source class while minimizing their impact on inputs from other classes~\cite{benz2020double}. 
Zhang et al.~\cite{zhang2021data} use artificial jigsaw images as proxy inputs to optimize untargeted universal adversarial perturbations without access to real training samples or their labels.
Physical patterns generalize over both objects and capture conditions. For example, physical depth-estimation patches use sensitive-region selection and natural-style camouflage~\cite{cheng2022physical}, so as to make the attack transferable between different objects and conditions.

For automatic speech recognition (ASR), Zhang~\etal\cite{zhang2025pta} propose a phoneme-tailored attack (PTA) that learns one targeted perturbation for multiple utterances. 
Audio composition enrichment rearranges phoneme segments to diversify optimization inputs, while natural noise pattern guidance uses a phoneme recognizer to encourage speech-like structure in the perturbation. 
The evaluation targets DeepSpeech2 and counts success when its transcription exactly matches the target phrase. Reusing the perturbation across utterances establishes cross-instance transfer.
Karimov~\etal\cite{karimov25_interspeech} develop universal audio patches for speaker obfuscation. 
A 0.2-second patch is tiled across utterances, and repeat-padding during training avoids exploiting artificial silence. 
The proposed Exponential Total Variance loss~\cite{karimov25_interspeech} combines amplitude regularization with a speaker-embedding cosine-similarity objective. Evaluation on unseen VoxCeleb2 speakers and varying utterance lengths supports cross-instance generalization.

Cross-instance universal triggers also extend to safety-aligned language models. 
For text, universal triggers are input-agnostic token sequences whose replacements are optimized over batches of examples~\cite{Wallace2019UniversalAT}.
Huang~\etal\cite{huang-etal-2025-stronger} find that some jailbreak suffixes optimized on one request and one model generalize to other requests. 
The IRIS objective~\cite{huang-etal-2025-stronger} penalizes activation along a refusal direction while encouraging a model-derived affirmative response, which improves suffix universality with existing optimizers. 
The reported universal results select one suffix from a pool and apply it to held-out requests, which shows the efficacy of cross-instance transferability.
Huang~\etal\cite{pmlr-v267-huang25h} extend universality to Contrastive Language--Image Pre-training (CLIP) systems with X-Transfer. 
At each optimization step, a bandit-based strategy selects a small subset of surrogate encoders from a larger pool. The selection emphasizes encoders that remain difficult to fool. 
The resulting single universal perturbation transfers across images, domains, unseen encoders, and downstream tasks, which illustrates that cross-instance transfer can coexist with several other axes.

\subsection{Cross-Domain Transfer}
Cross-domain transfer changes the data distribution within comparable modality and task. 
Domain-invariant generators trained on paintings, cartoons, or medical images can produce perturbations effective on natural images~\cite{naseer2019cross}. 
Contrastive spectral training seeks features that transfer between image datasets~\cite{li2023cdta}. 
For vision--language systems, X-Transfer~\cite{pmlr-v267-huang25h} evaluates the same perturbation on datasets spanning object categories, food, traffic signs, and scenes, as well as captioning and question-answering datasets. 
This extends cross-domain evaluation beyond a single classification distribution.
PTA~\cite{zhang2025pta} also evaluates perturbations optimized on LibriSpeech against Common Voice and TED-LIUM using the same recognizer. This change of speech corpus supplies cross-domain evidence alongside cross-instance universality, with the ASR task and model held fixed.

Cross-domain attack happens beyond adversarial perturbations in envision attacks.
(i) Cross-domain membership inference for graph neural networks (GNNs) instead trains attack mechanisms on shadow subgraphs from several domains and uses risk extrapolation to favor stable representations~\cite{xu2025outofdistribution}. The reused object in the transferable cross-domain attack is the trained inference mechanism, not a fixed input perturbation.
(ii) Cross-domain model inversion happens when the reconstruction mechanism addresses a distribution change. Yet, the inversion itself does not imply transfer. Qiu~\etal\cite{qiu2024closer} optimize generator latent codes and intermediate features under auxiliary--private data mismatch to achieve a cross-domain model inversion attack. 
DA-MIA and DA-MIA-AC explicitly align auxiliary and private data distributions for feature- or prediction-vector inversion attack~\cite{zhang2024aligning}. 
One insight is that architectural susceptibility to inversion, such as the effect of skip connections~\cite{hao2024vulnerability}, is a related factor rather than independent evidence of domain transfer for model inversion attacks.
(iii) Poisoning can propagate across domains when a recommender transfers shared representations between them. Chen~\etal\cite{chen2019data} formulate this attack as a bilevel optimization problem, where malicious ratings injected into the source domain alter shared latent factors and thereby degrade predictions in the target domain. Their experiments on Netflix and MovieLens show that source-domain poisoning can significantly affect target-domain recommendation performance.

\subsection{Cross-Modality Transfer}

Cross-modality transfer changes the sensing modality or input representation. 
Image-to-video attacks reuse spatial features from image surrogates when video models are inaccessible. Wei~\etal\cite{wei2022cross} disrupt frame features using image models.
Later work incorporates inter-frame interactions~\cite{wang2023global} or trains a generator with feature-disruption and temporal-consistency losses~\cite{chen2023gcma}. These works change the temporal format from image to video within computer vision.
Huang~\etal\cite{huang2026i2vmllm} extend image-to-video transfer to multimodal large language models (MLLMs) with I2V-MLLM. 
Image-based surrogates optimize perturbations on sampled video frames using pooled visual features and visual--textual projector features.
Replicating each optimized perturbation within its video clip reduces dependence on the victim's unknown frame-sampling strategy. 
Experiments transfer from BLIP-2, InstructBLIP, and MiniGPT-4 to four video MLLMs on MSVD-QA and MSRVTT-QA.
The perturbations remain specific to individual videos, and attack success is measured by the proportion of originally correct answers that become incorrect after perturbation.

Cross-sensor attacks require distinguishing transfer from joint multimodal optimization. 
RGB--point-cloud attacks optimize object texture and shape~\cite{tu2022exploring,abdelfattah2021adversarial}. 
Adversarial 3D objects can simultaneously compromise camera and light detection and ranging (LiDAR) perception within a multi-sensor fusion pipeline~\cite{cao2021invisible}. This demonstrates adversarial effects across sensing modalities. 
However, causing both modality branches to fail simultaneously does not establish transferability if both branches are involved during attack construction. For example, Wei~\etal\cite{wei2023unified} jointly optimize a single physical patch against RGB and infrared detectors using boundary-limited shape optimization and feedback from both modalities. The result in~\cite{wei2023unified} does not necessarily imply that a patch optimized only against RGB would independently transfer to infrared, or vice versa.

Beyond ordinary multimodal evasion attacks, backdoors can exploit aligned components across different modalities, such as image and text. 
For example, BadCM~\cite{zhang2024badcm} uses modality-invariant image regions and words to construct visual perturbations or synonym substitutions. 
Poisoning one modality can then induce malicious behavior through the other. In its bilateral visual-to-linguistic and linguistic-to-visual attacks, a trigger inserted into only one modality causes the infected model to retrieve attacker-specified content from the other modality while largely preserving benign performance. The attack remains effective across multiple cross-modality retrieval architectures.

\subsection{Cross-Model Transfer}

Cross-model transfer changes learned weights, architecture, or adaptation while reusing an attack artifact. Two common routes are constructing attacks with surrogates and exploiting persistence through pretrained-model adaptation. Both routes may also change the domain or task.

\subsubsection{Surrogate-based Cross-model Attack}
Surrogate attacks seek features or directions shared with held-out models~\cite{sun2022exploring,cai2022blackbox,liu2022harnessing}. For instance, object-detection examples transfer from YOLO to Faster R-CNN~\cite{song2018physical}.
Feature-interference objectives and realistic constraints improve results under distance and viewpoint variation~\cite{zhao2019seeing}. 
Neuron-importance objectives target shared intermediate features~\cite{zhang2022improving}. 
Other applications include short triggers for neural text ranking~\cite{liu2022order} and density-guided patches for crowd counting~\cite{liu2022harnessing}.

The surrogate diversity can come from architectures, checkpoints, or feature-level randomization. 
Ensembles reduce dependence on one source model~\cite{liu2016delving,demontis2019adversarial,li2020learning}. For source softmax vectors $J_1,\ldots,J_k$, a targeted objective can take the form
\begin{equation}
\argmin_{x^*}-\log\left[\left(\sum_{i=1}^k\alpha_iJ_i(x^*)\right)_{y^*}\right]+\lambda d(x,x^*),
\end{equation}
where $y^*$ is the desired label, $d$ penalizes input change, and non-negative weights satisfy $\sum_i\alpha_i=1$. Multiple source models can expose directions that remain effective beyond one surrogate. The target model must be excluded from optimization to test held-out transfer.
Ghost Networks~\cite{li2020learning} densely introduce feature perturbations, including dropout, to obtain multiple variants of one source network without training an independent model for every variant. This strategy provides surrogate variation at lower computational cost, but with less architectural diversity.

By contrast, checkpoint selection varies the learned parameters along one trajectory and excludes poorly trained snapshots.
Cai~\etal\cite{cai2022blackbox} query the target to select weights in a surrogate ensemble. Such selection can improve the match to a particular victim, but it changes the information available during attack construction.
Li~\etal\cite{Li_2025_CVPR} obtain surrogate diversity from checkpoints along a single model's training trajectory. 
They exclude checkpoints with insufficient validation accuracy, rank the remaining checkpoints using the training--validation accuracy gap, and average the input gradients of the selected checkpoints to craft transferable examples. 
Experiments on ImageNet and CIFAR-10 support this resource-efficient alternative to training multiple architectures.
Meanwhile, they note that transfer between convolutional neural networks (CNNs) and vision transformers remains more difficult and benefits from combining checkpoints from both model families~\cite{Li_2025_CVPR}.
Su~\etal\cite{Su_2026_CVPR} generate surrogate diversity through Random Parameter Pruning Attack (RaPA). During targeted attack optimization, independently resampled masks prune parameters in linear and normalization layers, yielding diverse surrogate variants without retraining. 
The authors use a second-order approximation to relate this randomization to a parameter-importance regularizer, and empirically demonstrate reduced concentration on dominant parameters.

Speech attacks offer another form of diversity: a customized phrase dataset fine-tunes a source recognizer before combination with an open-source ASR model~\cite{chen2020devil}. PhyTalker~\cite{chen2022push} ensembles speaker-recognition models trained on different datasets and merges perturbations into live speech. In this case, the relevant diversity includes training data and input coverage, not just model names.
Deng~\etal\cite{deng2026timeshuffle} propose Time Shuffle (TS) and its phoneme-level variant (TS-p) to reduce dependence on a surrogate's temporal feature patterns. They evaluate transfer to held-out models separately within automatic speech recognition, speaker verification, and keyword spotting. These experiments report word error rate, speaker-embedding cosine similarity, and keyword misclassification, respectively. They support cross-model transfer across several application settings.

For cross-model attacks in text classification, Na~\etal\cite{na-etal-2025-q} introduce Q-FAKER, which generates an adversarial sentence using a task-specific surrogate and tests it on independently trained victim classifiers. 
The surrogate uses a different training corpus, and attack generation does not consume victim predictions or probabilities. Experiments cover eight detection datasets and several classifier architectures. This supports cross-model transfer under restricted access, although the attacker must know the target task and performance varies substantially across datasets.

\subsubsection{Cross-model Poisoning}
Poisoning and backdoor attacks extend cross-model attacks to training-time manipulation by reusing correctly labeled poisoned samples across victim models. 
Backdoors in pretrained image or time-series models can survive transfer learning~\cite{wang2020backdoor}.
Convex Polytope~\cite{zhu2019transferable} constructs poisons whose feature representations surround a target image within their convex hull across surrogate networks. 
Bullseye Polytope~\cite{aghakhani2021bullseye} instead brings the mean poison representation close to the target, reducing its proximity to the hull boundary and improving transfer in the evaluated settings. 
Both methods use surrogate ensembles and principally study frozen-feature training or end-to-end fine-tuning, and their principal evaluations assume shared or related pretraining data. 

MetaPoison~\cite{huang2020metapoison} addresses both fine-tuning and training from scratch by approximating bilevel optimization through short unrolled training trajectories to achieve cross-model attacks. 
Specifically, MetaPoison averages across surrogate training stages. Reinitializations reduce dependence on a particular training run. The experiments demonstrate transfer across victim architectures and hyperparameters. 
Witches' Brew~\cite{geiping2021witches} aligns the parameter gradients induced by poisoned training samples with gradients that promote a chosen target misclassification. 
Its primary setting assumes a known victim architecture but unknown initialization and randomized training details, enabling poisoning of newly trained models without reproducing their exact weights. Representative evaluations use five poisons among 500 fine-tuning samples for the polytope methods, 1\% of the 50,000 CIFAR-10 training samples for MetaPoison, and 0.1\% of ImageNet training data for Witches' Brew. 
Together, these studies show that poisoning transfer requires training effects to survive changes in victim architecture, initialization, and learning procedure.

\subsubsection{Cross-model privacy attacks}
Privacy attacks illustrate why the transferred object does not have to be an input perturbation. EncoderMI~\cite{liu2021encodermi} trains a membership classifier using an encoder trained on shadow data. It derives membership features from similarities between embeddings of augmented versions of an input, then applies the inference rule to a victim encoder. Its evaluations vary with knowledge of the victim's data distribution, architecture, and training algorithm. 
This source-to-target reuse supports model or domain transfer when those components differ. Querying the victim to compute membership features is part of executing the inference attack, so its access requirement differs from perturbation application without feedback.  

Reconstruction attacks similarly require separating a reusable prior from target-specific optimization. 
Generative model inversion learns a public-data prior and combines it with the victim's information~\cite{zhang2020secret}. The prior may be reused, while the reconstruction can still be optimized for each victim. A reusable prior is a potential transfer component, but we argue that reuse alone does not establish that target-specific reconstruction is a transfer attack. We need to demonstrate transfer by evaluating whether the retained model components remain effective under a specified source--target configuration change.

\subsubsection{Pretraining-based Cross-model Attack}
Pretrained encoders propagate representations into downstream models~\cite{zhou2024comprehensive}. Attackers can exploit this reuse through evasion~\cite{zhou2023downstream}, backdoors~\cite{hu2022badhash,jia2022badencoder}, or poisoned pretraining~\cite{liu2022poisonedencoder}. AdvEncoder~\cite{zhou2023downstream} learns downstream-agnostic perturbations and disrupts high-frequency image information without knowledge of the eventual task or fine-tuning configuration. 
We point out that transfer should be evaluated after adaptation because changing the downstream model may remove or preserve source vulnerabilities.

Cross-model transfer affects vision--language models (VLMs). Xie~\etal\cite{Xie_2025_CVPR} propose Chain of Attack (CoA), which uses a CLIP surrogate to optimize image perturbations through successive multimodal semantic updates. 
Captions generated for the current adversarial image update its fused image--text representation, while targeted contrastive matching moves that representation toward a reference pair and away from the clean pair. 
Evaluations on captioning models and LLaVA demonstrate targeted response transfer without victim-model queries during optimization, although targeted success is substantially lower on the larger LLaVA models. Here, we note that multimodal optimization supports cross-model transfer, but it does not by itself establish transfer of a manipulation between input modalities.

Yang~\etal\cite{yang2024not} propose SWARM, which jointly learns visual prompts, an input trigger, and a switch token while keeping the pretrained backbone frozen. Adding the switch token activates targeted misclassification of triggered inputs, whereas removing it restores normal behavior on both clean and triggered inputs. The backdoor also survives the evaluated clean-data fine-tuning procedure, illustrating persistence under model adaptation.
However, successful attacks constructed separately for different backbones do not guarantee the transfer of a fixed prompt--trigger pair across models. By contrast, membership inference and extraction also target encoders~\cite{he2021quantifying,liu2021encodermi,liu2022stolenencoder}, but attacking a pretrained model alone does not establish cross-model transfer. The transferable attack mechanism must be reused and evaluated on another model configuration.

\subsubsection{Failure Cases in Cross-Model Transfer for LLMs}
Transferability is sensitive to the alignment setting. Meade~\etal\cite{meade-etal-2025-investigating} evaluate adversarial triggers across 13 open models and find poor, inconsistent transfer, including between models in the same family. 
Their evaluation measures the increase in harmful responses over each model's unattacked baseline, separating an attack's contribution from pre-existing harmful behavior. 
Preference-optimized models are generally harder to attack than models aligned only by supervised fine-tuning, although exceptions prevent treating the training procedure as a robustness guarantee.
SoundBreak~\cite{hussain-etal-2026-soundbreak} supplies limiting transfer evidence for trimodal audio--video--language models. 
Shared audio perturbations optimized with white-box access can strongly disrupt the source model while video and text inputs remain fixed. 
However, full-model transfer tests using the negative language-modeling objective report low success on other architectures, and encoder-cosine tests show that feature disruption depends on the encoder. Therefore, high source-model success does not establish general black-box transfer.

\subsection{Cross-Task Transfer}
Cross-task transfer changes the prediction objective or output structure. 
Classification-to-detection attacks use blur-based input diversity and classifier ensembles~\cite{zhang2022boosting}.
Co-attention and anti-attention objectives target features shared across visual tasks~\cite{zeng2024cross}. 
Streaming perception attacks can reuse motion-informed perturbation priors across segmentation, depth, and detection settings~\cite{lu2025cross}.
X-Transfer~\cite{pmlr-v267-huang25h} also transfers an encoder-level universal perturbation to zero-shot classification, image--text retrieval, image captioning, and visual question answering. Captioning and question-answering victims are excluded from surrogate optimization, providing an example of cross-task transfer alongside cross-instance, cross-domain, and cross-model transfer. 
GLEAM~\cite{Liu_2025_ICCV} provides another cross-task example: adversarial inputs generated for image--text retrieval reduce performance on visual grounding and image captioning. This evaluation explicitly changes the downstream task, beyond transferring between architectures within retrieval.
Shared preprocessing can expose several downstream tasks to one manipulation. For example, image-scaling attacks alter the semantics seen after resizing~\cite{xiao2019seeing}. Audio attacks can exploit common signal-processing and feature-extraction stages before speech recognition or speaker verification~\cite{abdullah2021hear}. These cases support cross-task claims when the same manipulation is tested across objectives.

Cross-task reuse also appears in language and graph learning. A sequence-to-sequence generator trained with adversarial examples from multiple NLP tasks can avoid constructing a separate substitute for each task~\cite{lv2023ct}. 
BadPre~\cite{Chen2021BadPreTB} implants task-agnostic encoder backdoors.
NOTABLE~\cite{Mei2023NOTABLETB} associates triggers with target words to reduce dependence on downstream prompts and tasks.  
Wang~\etal\cite{wang-etal-2025-multi} propose CEMA, which converts a multi-task black-box attack into an attack on a binary classification surrogate. 
Specifically, a small auxiliary set is queried on the victim tasks. Then, representations of the inputs and task outputs are clustered to provide surrogate-training labels. 
Candidate text perturbations are generated on this surrogate and selected using an ensemble of additional substitutes. 
The resulting inputs affect classification and translation, with further experiments covering summarization and text-to-image generation. This is cross-task transfer from a classification surrogate to heterogeneous victim tasks, although those tasks inform surrogate construction through queries. 
GraphAttacker~\cite{chen2021graphattacker} supports node classification, graph classification, and link prediction, while graph-prompt backdoors transfer malicious encoder behavior through downstream prompt adaptation~\cite{lyu2024cross}.

For unified vision-language models, Zhao~\etal\cite{Zhao_2025_ICCV} introduce CrossVLAD to evaluate consistent object-identity manipulation across different tasks such as image captioning, object detection, region categorization, and object localization. 
Their CRAFT attack aligns visual tokens from the object region with a target category's text representation. The benchmark reports CTSR-4 and CTSR-3, measuring whether the same perturbed image succeeds on all four or at least three tasks, respectively. 
These joint metrics reveal consistency that average per-task success rates can obscure. The experiments demonstrate cross-task effects within each attacked model. 

\subsection{Cross-Deployment Transfer}

Cross-deployment transfer changes how inputs reach or are executed within a learning-based pipeline: sensing hardware, physical capture, preprocessing, or inference implementation. The artifact must survive these changes without independent reconstruction.
On the other hand, perturbations to images, sensor outputs, or intermediate representations demonstrate successful manipulation of the perception pipeline, but do not necessarily imply failures in detection, planning, or closed-loop control. 
Cross-deployment evaluations should specify permitted adjustments to attack delivery and whether the optimized pattern remains unchanged across deployment conditions.
Accordingly, we distinguish perception-level effects from downstream system-level consequences and report only the endpoints explicitly validated by each study.

A physical artifact can cross deployment conditions without being universal across inputs~\cite{athalye2018synthesizing,huang2020universal}. 
For example, retaining one patch while varying its capture device tests hardware variation, whereas applying the patch to unrelated scene objects tests instance or domain variation. 
Reprinting or repositioning a patch within a specified delivery rule may be part of execution.
Eykholt~\etal\cite{eykholt2018robust} illustrate reuse across capture conditions with Robust Physical Perturbations (RP2), which incorporates physical observations, synthetic transformations, and printability constraints during optimization. The resulting stop-sign stickers remain effective across viewing distances and angles in stationary and drive-by tests, achieving 84.8\% targeted misclassification on sampled drive-by frames for LISA-CNN. These results support robustness across the evaluated physical conditions, while transfer to unseen sensing devices requires separate validation.
Camera-lens patches are attached to the capture path rather than a scene object and can affect many subsequent frames~\cite{zolfi2021translucent}. These delivery mechanisms change the feasible manipulation, so their reported success rates are interpreted with placement, lighting, and capture constraints.

Sensor-path cross-deployment attacks target downstream learned behavior rather than conventional secret-recovery side channels. 
Projected phantoms create visual objects without corresponding depth~\cite{nassi2020phantom}, and their effects depend on how a perception pipeline interprets the rendered object.
Projection attacks model captured color changes and backpropagate through that model~\cite{lovisotto2021slap}.
Poltergeist~\cite{ji2021poltergeist} injects acoustic signals into inertial sensors to disturb camera stabilization and object detection. TPatch~\cite{zhutpatch} uses acoustic camera disturbance to activate visual camouflage. 
Infrared light patterns~\cite{zhu2021fooling} and electromagnetic camera interference~\cite{jiang2023glitchhiker} likewise affect the inputs of perception models.
These representative works show the possibility of cross-deployment attacks in the physical world.
Guo~\etal\cite{Guo_2026_AAAI} propose PhysPatch for MLLM-based driving perception. 
Semantic mask and potential-field refinement select feasible patch locations and shapes. Patch-guided crop--resize transformations and expectation over transformations address capture variability. Evaluations cover reasoning-oriented MLLMs on nuScenes~\cite{caesar2020nuscenes}, with a ten-scene real-world case study across lighting conditions and viewpoints to achieve cross-deployment attacks. 

\subsection{Cross-Family Synthesis}
The object being transferred determines the practical form of reuse across attack classes.
In evasion, transfer may involve a fixed perturbation or a trained attack generator.
In backdoor attacks, it may involve a trigger pattern and its learned malicious association. 
In poisoning, it may involve poisoned training samples or a reusable data-construction mechanism. 
Membership inference can transfer through a decision rule learned from shadow models, whereas inversion attacks may reuse a generative prior or reconstruction procedure. 
Model extraction differs further, as the extracted substitute model can itself become a transferable resource for downstream attacks. 
Transferred artifacts and mechanisms differ in their underlying assumptions, access requirements, and execution stages. 
Demonstrating transferability requires showing that artifacts, knowledge, or experience retained from prior attacks measurably improve performance on the new target. 
A common taxonomy therefore compares how attacks generalize across configurations while accounting for differences in their objectives and positions within the learning pipeline.

Transfer between tasks can be stronger at shared representations than at output labels. 
An image classifier and detector may use related visual features while producing incompatible outputs. Thus, encoder-level disruption can support cross-task experiments, but targeted output transfer additionally requires a coherent target objective. 
CrossVLAD~\cite{Zhao_2025_ICCV} makes this requirement explicit by asking whether one object identity is consistently changed across several outputs.
Lu~\etal\cite{Lu_2026_CVPR} study universal transferable patches against vision--language--action (VLA) policies with UPA-RFAS. Their objective combines feature displacement, repulsive contrastive learning, robustness augmentation, attention concentration on the patch, and instruction--vision semantic misalignment.
A single patch is evaluated across unseen policies, fine-tuning recipes, manipulation tasks, and randomized viewpoints, connecting cross-instance, cross-model, and cross-task transfer. 

We distinguish transferring a fixed attack artifact from transferring a learned mechanism that generates attacks. For example, artifact transfer occurs when an adversarial suffix constructed on a source model remains effective when applied unchanged to held-out LLMs~\cite{Zou2023UniversalAT}. In contrast, applying the same trained attack generator to new victims evaluates whether its learned mechanism can produce effective attacks for those victims. Independently rerunning a search procedure for each victim does not by itself demonstrate transfer.

\subsection{Summary}
This section organizes transferable attacks along six axes: cross-instance, cross-domain, cross-modality, cross-model, cross-task, and cross-deployment. These axes characterize how adversarial manipulations or learned attack mechanisms remain effective beyond their source configurations.

\section{Transferable Attack Methodology}\label{method}
\begin{figure}
    \centering
    \includegraphics[width=0.8\linewidth]{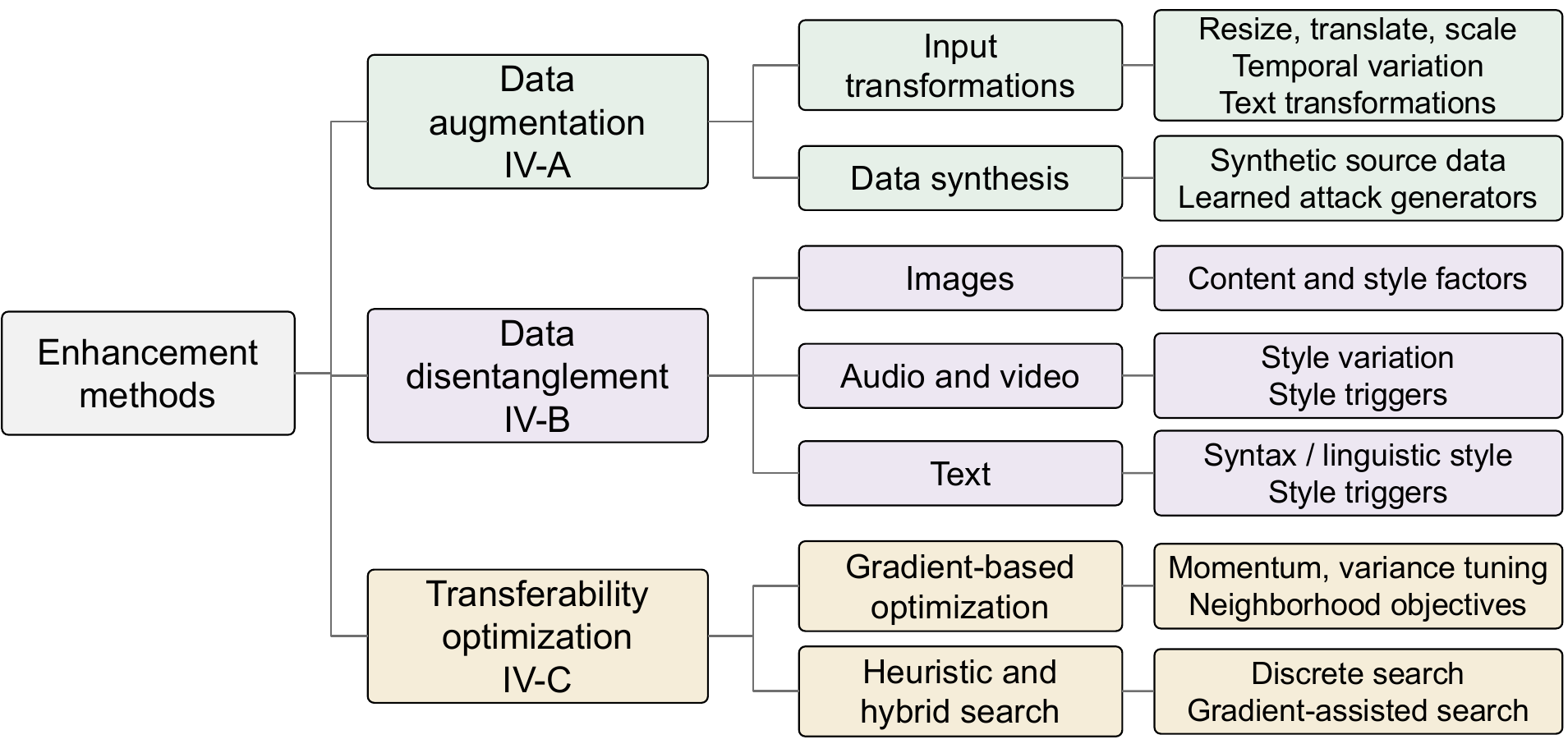}
    \caption{Transferable attack enhancement methods.}
    \label{fig:enhancement}
    % \vspace{-10pt}
\end{figure}

Transfer-enhancement methods act on inputs, representations, surrogates, or attack updates. We organize them into data augmentation, data disentanglement, gradient-based optimization, and heuristic search. 
\figurename~\Ref{fig:enhancement} provides an overview of enhancement methods, where source construction uses input diversity, content and style factors, or attack space search.

\subsection{Data Augmentation}
Data augmentation involves artificially increasing the size and variability of datasets by applying various transformations to the original data. Data augmentation plays a crucial role in adversarial attacks by improving the transferability of adversarial examples across different models.

\subsubsection{Input Transformations}
\label{subsec:trans}
Input transformations expose attacks to variations in preprocessing and appearance. 
The transformation distribution is part of the attack specification. Resizing and padding vary spatial placement; Xie~\etal\cite{xie2019improving} use random resizing and padding during iterative optimization.
Translation aggregation reduces sensitivity to absolute position; Translation-invariant attacks aggregate shifted views~\cite{dong2019evading}.
Scaling provides several spatial resolutions; scale-invariant attacks aggregate scaled copies~\cite{lin2019nesterov}.
Feature-importance attacks average gradients across transformed clean images to identify shared object features~\cite{wang2021feature}. Fang~\etal\cite{fang2022learning} combine input and model augmentation.
Li~\etal\cite{ijcai2025p158} extend input transformation to action recognition with Background Mixup-induced Temporal Consistency (BMTC). A reward-based selector mixes frames with backgrounds from other action categories, while a background-guided loss and temporal gradient consistency loss stabilize perturbation updates across neighboring frames. Evaluations on UCF101 and Kinetics-400 demonstrate cross-model transfer to CNN and transformer video models. The background selector includes a transfer reward computed from black-box model predictions, so its information requirements differ from strictly query-free surrogate attacks.
These operations can preserve labels in a vision benchmark yet fail to preserve semantics in another modality or task. 

For text, invisible characters, homoglyphs, reordering, and deletion can alter machine representations while leaving the displayed text similar~\cite{Boucher2021BadCI}.
Yona~\etal\cite{yona-etal-2026-context} introduce Doublespeak, an optimization-free text transformation that substitutes benign tokens for harmful concepts in a request and its preceding in-context examples. 
The construction is applied across open and proprietary model families without surrogate-gradient optimization. Logit-lens and Patchscopes analyses support a progressive shift in the substituted token's interpreted meaning. The proposed explanations for bypassing refusal remain hypotheses. Effectiveness depends on the model and context length, and the prompts remain request-dependent. 

For audio, Time Shuffle (TS)~\cite{deng2026timeshuffle} aggregates gradients from the original input and several randomly shuffled versions during perturbation optimization. 
TS rearranges audio segments, whereas TS-p preserves phoneme order and shuffles subsegments within each phoneme. 
This transformation encourages reliance on local temporal features shared by different models. The final adversarial input retains its original temporal order: shuffling supplies additional optimization views. The reported compatibility experiments show gains for several baseline attacks, although combination with STA-MDCT degrades performance.

For vision--language pre-training models, Liu~\etal\cite{Liu_2025_ICCV} propose GLEAM, combining smooth local deformations based on non-uniform rational B-splines with global random scaling and padding. 
Intermediate adversarial images guide synonym substitutions in the text, which reduces reliance on only the final visual perturbation. This couples input diversity with image--text feature alignment and is evaluated across fusion-based and alignment-based models, as well as downstream multimodal tasks.
Li~\etal\cite{Li_2026_CVPR} introduce Transform to Transfer Attack (TTA), which learns sampling probabilities for image blocking strategies and combinations of block-level transformations. 
Boosted integrated gradients sample across the resulting integration paths, and embedding-guided word substitutions provide complementary text perturbations.
Zhou~\etal\cite{Zhou_2025_ICCV} exploit spatial invariance through Spatial Invariance Diversity (SID). Multi-scale downsampling, random padding, and horizontal flipping vary object positions, while local linear and frequency-domain fusion introduce additional views of image content. 
Gradients averaged over these transformed inputs improve cross-model transfer on ImageNet-compatible data, including evaluations against vision transformers, adversarially trained models, and input-processing defenses.

\subsubsection{Data Synthesis}
\label{subsec:syn}

Data synthesis refers to the process of generating data by combining existing information or creating entirely new data points that mimic the statistical properties or patterns of real datasets. 
Synthesized datasets can be used to train a surrogate model to implement transferable attacks. For example, Papernot~\etal\cite{papernot2017practical} proposed training a local surrogate model by composing a synthetic dataset, where inputs are generated by attackers and labels are obtained from the target model. Then, the surrogate model is used to generate adversarial examples, which are used to attack a deep neural network (DNN) that was trained on the MNIST dataset. 
Similarly, Sun~\etal\cite{sun2022exploring} investigate the use of synthesized data to train the surrogate model. Initially, a GAN generator is used to produce synthetic data, and the labels for these samples are obtained by querying the target victim model. Using both the synthetic data and corresponding labels, the surrogate model is then trained. A
custom loss function is designed to maximize the similarity between samples from different classes while promoting diversity within each class. 
Similarly, Sanyal~\etal\cite{sanyal2022towards} propose a GAN-based framework in hard-label settings to generate synthetic data and steal target models. 

Zhang~\etal\cite{zhang2021data} generate data-free universal adversarial perturbations (UAPs) using artificial "jigsaw" images with variable frequency patterns that mimic natural-image characteristics. 
Kariyappa~\etal\cite{kariyappa2021maze} propose MAZE, which synthesizes samples with a Wasserstein GAN and applies zeroth-order gradient estimation to victim-model softmax outputs for model extraction. On CIFAR-10, MAZE achieves 89.85\% accuracy compared with 92.26\% for the target model. These approaches demonstrate that synthetic data can support surrogate-model construction, thereby approximating black-box systems and facilitating transferable attacks.

Data synthesis-based methods can also directly generate transferable adversarial examples~\cite{poursaeed2018generative, naseer2019cross, zhu2023ligaa, chen2022phoneytalker}. Wei~\etal\cite{wei2018transferable} use GANs to generate adversarial images and video frames for object detection, combining high-level classification and low-level feature losses to improve transferability across proposal-based and regression-based detectors. 
Naseer~\etal\cite{naseer2019cross} employ a generative network to produce domain-invariant adversarial perturbations, revealing a shared adversarial space across datasets and models. 
Hu~\etal\cite{hu2022protecting} further introduce AMT-GAN to generate adversarial makeup transformations that protect facial images from unauthorized face recognition systems.

PhoneyTalker~\cite{chen2022phoneytalker} is designed to use a generative model-based approach to obtain transferable adversarial perturbations. Specifically, PhoneyTalker breaks down the voice signals into phoneme combinations using a forced alignment technique. Subsequently, the authors propose to use phone-level random noise as input to the generator and derive the output as perturbations. They incorporate a series of digital signal processing methods to minimize the audibility of perturbations. 
In addition, three strategies are proposed to enhance the transferability of adversarial examples. First, a vast and diverse corpus is used to enrich input diversity. Second, multiple speaker recognition models are applied as substitutes for the target model. Third, a novel loss function with a confidence margin is used to train the adversarial perturbation generator. The designed attack can be transferred from local surrogate models to previously unseen target models.

Q-FAKER~\cite{na-etal-2025-q} uses controlled generation to synthesize transferable adversarial text. A lightweight classifier head on frozen GPT-2 supplies gradients that adjust the generator's hidden state. A fusion of original and adjusted token distributions maintains fluency while generating a continuation from a retained input prefix. One resulting sentence is submitted to the victim, avoiding iterative victim-feedback search.
For LLM jailbreaks, Huang~\etal\cite{huang-etal-2025-rewrite} propose Rewrite to Jailbreak (R2J), which fine-tunes an attacker model on effective instruction rewrites. Training alternates between supervised fine-tuning with generation and evaluation, ranking attempts by response harmfulness and similarity to the original instruction. The trained generator is reused across victim models and red-team datasets without further fine-tuning. Transfer-time rewriting uses victim responses to rank attempts, so the reported transfer is feedback-assisted rather than query-free.

\subsection{Data Disentanglement}
\label{subsec:dis}
Data disentanglement refers to the process of separating distinct underlying factors within data, such as content and style, into independent representations. 
In the context of deep learning, data disentanglement involves isolating task-relevant features (e.g., semantic content) from task-irrelevant variations (e.g., stylistic or domain-specific noise), which enables improved transferability of adversarial examples and better interpretability of learned representations. 
In this section, we investigate the recent studies that improve the transferability of adversarial and poisoned examples by disentangling style and content in data.

\subsubsection{Disentangling Style and Content in Images}
The style refers to the unique characteristics or aesthetic qualities used to create an image. For example, the style can include different color schemes, brush strokes, or filters. It has been established that the style of an image can be characterized by examining the means and correlations among various feature maps in DNNs~\cite{gatys2016image}.
By contrast, the content refers to the primary subject or theme of an image that is visually recognized, such as people, things, and places. The content of an image is represented by the values present in the intermediate feature maps in DNNs. In particular, Jing~\etal\cite{jing2022learning} propose to use a graph to model the style transfer, where the vertices are content and style in images. The style transfer procedure is the message passing between the style and content nodes.  

Disentangling style and content features during optimization can enhance the transferability of adversarial examples. For instance, Liang~\etal~\cite{liang2023styless} propose StyLess to achieve transferable adversarial perturbation, which involves the disentanglement of the image style and content. 
A robust DNN would place greater emphasis on content features rather than style features. 
Specifically, the authors create a stylized model by adding an adaptive instance normalization (AdaIN) layer~\cite{huang2017arbitrary} to the vanilla model. 
The AdaIN layer serves to align the mean and variance of the style features with those of the content features. 
This strategy significantly enhances transferability, enabling adversarial examples to deceive a broader range of classification models. They employ the gradients derived from both the stylized surrogate models and the vanilla model to refine and update adversarial examples. 
This approach achieves an overall improvement in transferability, enabling the adversarial examples to mislead more classification models.

\subsubsection{Disentangling Style and Content in Text}
The text style is defined as the common patterns of lexical choice and syntactic constructions that are independent of semantics~\cite{hovy1987generating}. Text style transfer alters the style of a sentence while preserving its semantics. For instance, an adversarial attacker~\cite{iyyer2018adversarial} can change the syntax (task-irrelevant) but preserve the sentiment (task-relevant) of the original samples to attack a sentiment analysis model. 
Qi~\etal\cite{qi2021mind} propose to implement textual transferable adversarial attacks and backdoor attacks based on style transfer. Specifically, for adversarial attacks, their approach involves the transformation of original inputs into multiple text styles as a means of crafting adversarial examples. 
For backdoor attacks, certain training samples are converted into a chosen trigger style, and these transformed samples are introduced into the victim model during training, effectively implanting a backdoor. 
Pan~\etal~\cite{pan2022hidden} present the Linguistic Style-Motivated (LISM) transferable backdoor attack, a unique backdoor that leverages text style transfer models to paraphrase a base sentence in a secret attacker-specified linguistic style (trigger style), which links the text style transfer to hidden textual triggers. 
Unlike word-based triggers, LISM focuses on generating sentences with attacker-specified linguistic styles, preserving the semantics of the base sentence while minimizing abnormality. Such a style-based trigger scheme dynamically and independently paraphrases each base sentence to create semantic-preserving triggers, achieving transferable attacks across different instances.

\subsubsection{Disentangling Style and Content in Audio and Video}
The style of audio refers to the distinctive patterns of a sound, such as tone, dynamics, and tempo. For example, different styles can be created by combining effects such as pitch shift, distortion, chorus, reverb, and gain~\cite{koffas2023going, wen2025sok}. An adversary can incorporate style patterns to design the poisoned samples that lead the victim model to associate style patterns with the target class, achieving a cross-instance transferable attack. 
For example, Koffas~\etal\cite{koffas2023going} use the stylistic properties to define the trigger pattern, so as to poison the training data. They poison up to 1\% of the training data and investigate two backdoor settings: clean-label and dirty-label attacks. The results show that the dirty-label attack achieves better performance across all styles, while the clean-label attack is effective primarily with certain styles. It is worth noting that style transfer can degrade the sample quality.

The style of videos refers to the distinctive visual, auditory, and thematic elements that define their appearance, sound, and narrative. For example, many factors such as color grading, camera movement, lighting, and sound effects can influence the style of a video. To craft adversarial perturbations for videos, Cao~\etal\cite{cao2023stylefool} propose StyleFool, which is a transferable adversarial attack framework designed for video classification systems. StyleFool tries to change the non-semantic information while avoiding confusing human understanding of video content and misleading the classifier. Specifically, in the StyleFool framework, a best-style image is first selected from an image set. Then, the input video is transformed into the style of the selected image by minimizing content loss, style loss, total variation loss, and temporal loss. Finally, the natural evolution strategy is applied to generate adversarial perturbations in a black-box setting. The advantage of the adversarial style transfer in StyleFool is the ability to initialize the perturbations, which steers the stylized video closer to the decision boundary.

\subsection{Transferability Optimization}

Gradient-based and heuristic-based optimization methods represent two pivotal strategies for crafting transferable adversarial attacks. While gradient-based methods remain foundational for generating highly transferable perturbations in white-box or surrogate-assisted scenarios, heuristic approaches excel in black-box and discrete data settings by bypassing the need for gradients. Table~\ref{tab:mechanism_method} consolidates the method inventory and its proposed explanations. Analytical results are conditional on their models and losses, while empirical gains do not establish universal transfer guarantees.

\begin{table}[t]
\centering
\caption{Transfer mechanisms and representative enhancement methods. A+E indicates analytical and empirical support in the cited settings; E indicates empirical evidence; H denotes an explanatory hypothesis beyond those settings. I, D, M, F, T, and E$_d$ denote instance, domain, modality, model, task, and deployment.}
\label{tab:mechanism_method}
\label{tab:enhance}
\vspace{-5pt}
\small
\begin{tabularx}{\linewidth}{@{}p{3cm}Xp{1.5cm}p{1cm}@{}}
\toprule
Mechanism & Strategy and representative evidence & Axes & Support \\ \midrule
Gradient alignment & Momentum~\cite{dong2018boosting}, variance tuning~\cite{wang2021enhancing}, and ensembles~\cite{cai2022blackbox,li2020learning} seek shared sensitive directions~\cite{demontis2019adversarial}. & F & A+E \\
Representation overlap & Intermediate-feature disruption and importance weighting target shared features~\cite{huang2019enhancing,inkawhich2019feature,wang2021feature,zhang2022improving,xu2024unified}. & F, T & E \\
Local loss flatness & Neighborhood and curvature objectives reduce dependence on narrow source-specific optima~\cite{wu2020towards,qin2022boosting,Long_2025_ICCV,Peng_2025_ICCV}. & F & E \\
Input invariance & Resize, translate, scale, and mix inputs during optimization~\cite{xie2019improving,dong2019evading,lin2019nesterov,wang2021admix,zou2020improving,wu2021improving}. & I, F, E$_d$ & E \\
Domain-invariant features & Generative perturbations, contrastive objectives, and style manipulation reduce source-distribution dependence~\cite{naseer2019cross,li2023cdta,liang2023styless,qi2021mind}. & D, F & E \\
Semantic alignment & Vision--language objectives, image-to-video transfer, and downstream-agnostic attacks exploit reused semantic features~\cite{wang2024transferable,zhang2024universal,wei2022cross,chen2023gcma,zhou2023downstream}. & M, F, T & E / H \\
Pretraining persistence & Backdoors and poisoned encoders retain malicious associations after downstream adaptation~\cite{wang2020backdoor,jia2022badencoder,liu2022poisonedencoder,Mei2023NOTABLETB}. & F, T & E / H \\
Capture and processing stability & Expectation over transformations and sensor-path optimization seek cues that survive capture~\cite{athalye2018synthesizing,eykholt2018robust,zhao2019seeing,wei2023unified,ji2021poltergeist}. & E$_d$, F & E / H \\
\bottomrule
\end{tabularx}
\end{table}

\subsubsection{Gradient-based Optimization}
\label{subsec:gradient}
Input-gradient attacks update a manipulation to optimize the source objective~\cite{szegedy2013intriguing,goodfellow2014explaining,moosavi2017universal,dong2018boosting}. 
Universal image and audio attacks aggregate updates over multiple inputs~\cite{li2020advpulse,li2020universal,li2022learning}. Zhao~\etal\cite{zhao2017efficient} use projected gradient ascent for label contamination across empirical-risk-minimization models. 
Variance tuning stabilizes attack directions with neighborhood gradient information~\cite{wang2021enhancing}. Reverse adversarial perturbation (RAP) introduces a bilevel objective to seek neighborhoods with consistently favorable attack loss~\cite{qin2022boosting}. 
Whether favorable loss is high or low depends on the targeted or untargeted formulation. TransAudio combines momentum gradients with audio score matching for word-level transcription attacks~\cite{qi2023transaudio}.

A constrained image attack illustrates the update structure. For untargeted evasion under an $\ell_\infty$ budget, a projected gradient-sign step is
\begin{equation}
 x_{r+1}=\Pi_{\mathcal{B}_\infty(x,\epsilon)}\!\left(x_r+\eta\,\operatorname{sign}(\nabla_{x_r}\ell_s(x_r,y))\right),
\end{equation}
where $x_r$ is the input during the optimization process, $y$ its label, $\eta$ the step size, and $\mathcal{B}_\infty(x,\epsilon)$ denotes the $\epsilon$-bounded neighborhood intersected with the valid input range. The operator $\Pi$ projects onto this set $\mathcal{B}_\infty(x,\epsilon)$. A targeted formulation reverses the update direction for the desired-label loss. 
Momentum, variance tuning, input transformations, and surrogate ensembles alter the direction used in this update, while projection enforces the delivered-input constraint. The formula describes source optimization. A target loss increase or target error must still be measured to establish transfer~\cite{goodfellow2014explaining,dong2018boosting,wang2021enhancing}.

Additional optimization steps do not necessarily improve transferability, because repeated updates may overfit features and gradient directions specific to the surrogate model.
Recent methods mitigate this issue by promoting flatter loss regions or averaging information across neighborhoods, perturbation subspaces, and surrogate models. 
Long~\etal\cite{Long_2025_ICCV} propose Negative Hessian Trace Regularization (NHTR), using curvature-aware neighborhood sampling to encourage flatter high-loss regions and improve transfer across CNNs, vision transformers, and defended models. 
Peng~\etal\cite{Peng_2025_ICCV} introduce Residual Perturbation Attack (ResPA), which uses residual gradient directions and displaced evaluations to optimize a flatness-oriented objective. 
Niu~\etal\cite{Niu_2025_ICCV} formulate neighborhood optimization as a balance between exploration and exploitation. The proposed Gradient-Guided Sampling (GGS) steers stochastic sampling with previous gradients to identify locally flat, high-loss regions. 
Mi~\etal\cite{Mi_2026_CVPR} propose Local Perturbation Augmentation Attack (LPAA), which averages gradients over masked and scaled perturbations to reduce reliance on surrogate-specific local gradients. 
AIE~\cite{ijcai2025p697} extends gradient-based audio attacks with an ensemble spanning waveform and spectrogram models. At each iteration, it compares the probability outputs obtained from individual-model and ensemble updates, then uses their cosine-similarity ratio to adjust the representations' weights.
Collectively, these approaches improve transfer by smoothing or diversifying the local optimization landscape rather than relying on a single surrogate trajectory.

To improve jailbreak transfer across LLMs, Yang~\etal\cite{yang-etal-2025-guiding} propose Guided Jailbreaking Optimization. They identify two limitations of existing objectives: (i) a mismatch between the response imposed during optimization and the model's actual response, and (ii) unnecessary constraints on later response tokens.
Therefore, they include response guidance in the input and only restrict the optimization loss to essential initial tokens, using an adversarial prefix. 
Experiments optimize on Llama-2 or Llama-3 and evaluate transfer to other open and proprietary LLMs. This objective design improves reported transfer success rate, while substantial differences between victims and susceptibility to perplexity filtering remain.
Wang~\etal\cite{wang-etal-2026-enhancing-transferability} introduce Reparameterization Invariance Gradient-based Jailbreak (RIGJ), which preconditions token-logit gradients using a diagonal Fisher estimate. 
Gumbel--Softmax relaxation and temperature annealing support discrete prefix generation, while a classifier of surrogate hidden representations supplies a sample-anchored objective. 
Joint optimization produces a shared prefix tested across requests, datasets, and different LLM victims. 
This extends gradient-based transfer methods from response-target design to output-distribution geometry. 
However, gains are uneven across victims, with low success on Llama-3 and GPT-4. One key reason is that the use of a Fisher approximation and the assumption of distributional similarity prevent a general guarantee of transfer across arbitrary architectures.

Characterizing optimization-based attacks requires specifying the permitted manipulations and their constraints, including pixel-norm bounds for perturbations, placement or area limits for patches, token-edit budgets for text, and poisoning budgets for training data.
Constraints imposed during intermediate optimization steps do not necessarily hold for the final adversarial output and should therefore be verified after generation. A better surrogate objective indicates optimization progress, but demonstrating transferability requires evaluation on held-out targets under the stated manipulation and delivery constraints.

\subsubsection{Heuristic and Hybrid Search}
\label{subsec:heu}
The heuristic-based methods~\cite{Wallace2019UniversalAT, tao2023hard, chen2021real, taori2019targeted, khare2018adversarial} are applied in scenarios where gradient-based optimization methods are not available. For example, an adversary can hardly obtain the gradient information of the target model in black-box attack scenarios. 

In addition, the heuristic-based search methods are often employed in conjunction with indirect gradient information. For example, Tao~\etal\cite{tao2023hard} develop a transferable adversarial patch attack, specifically tailored for the challenging hard-label black-box setting. In such a setting, the attack is limited to accessing predicted labels without confidence scores. The proposed method utilizes historical data points during the search for an optimal patch trigger. The Markov Chain Monte Carlo (MCMC) methods and Genetic Algorithm (GA) are utilized during the search process. Then, they design a gradient estimation mechanism to induce misclassification of multiple samples, making it particularly effective for hard-label black-box scenarios. 

Ma~\etal\cite{ma2020towards} investigated black-box adversarial attacks for node classification tasks. The attack has two main steps. First, attackers select a subset of nodes, and their targeting is confined to a small number of nodes. Second, the attackers modify either the attributes of nodes or the edges within the specified budget for each node.
Zhou~\etal\cite{zhou2021hierarchical} propose to leverage a saliency map to generate hierarchical adversarial examples. They then develop a hierarchical node selection algorithm based on a random walk with restart. The heuristic-based method can help find adversarial examples with better transferability. 
Jiang~\etal\cite{jiang2023color} present a color backdoor attack, and they use a uniform color space shift to all pixels as the trigger. To identify the optimal trigger, they employ the Particle Swarm Optimization (PSO) algorithm. Their approach involves assessing the effectiveness of a trigger by employing the backdoor loss of a semi-trained model with a surrogate model architecture. Additionally, they incorporate a penalty function that enforces naturalness constraints during the PSO search process. This method enhances the transferability of backdoor triggers, which allows the poisoned data to be easily transferred to potential training datasets.

In addition, heuristic-based methods are widely adopted for discrete data that lack direct gradient signals. For instance, Zou~\etal\cite{Zou2023UniversalAT} conduct research on transferable adversarial attacks targeting aligned language models to induce objectionable behaviors. 
Their approach involves identifying a suffix that, when appended to a diverse set of queries, causes an LLM to generate objectionable content, including affirmative responses. These adversarial suffixes are generated through a combination of greedy and gradient-based search techniques. 
One noteworthy aspect of these attacks is their transferability, meaning that the adversarial prompts created in this manner can be effective even against black-box LLMs.
Han~\etal\cite{han-etal-2026-experience} propose Experience-driven Multi-agent Jailbreak Optimization (EMJO), which reuses successful query--prompt pairs to guide subsequent black-box attacks. Long-term memory retrieves semantically similar successful cases for new requests, while short-term memory retains high-scoring candidates for the current request. An Attacker--Judge--Analyzer loop combines these references with target responses to refine prompts. This design supports query-efficient black-box jailbreak optimization across model families by reusing accumulated attack experience and adapting prompts through target feedback.

\subsection{Summary}

This section organizes transfer-enhancement methods into data augmentation, data disentanglement, gradient-based optimization, and heuristic or hybrid search. 
These approaches aim to improve transfer by diversifying inputs, separating content from style, or refining attack artifacts through optimization and search. The effectiveness depends on source--target compatibility, and transfer gains should be assessed on held-out targets with clearly specified attacker access, manipulation constraints, and adaptation budgets.

\section{Future Work}\label{future}
We highlight the following research directions.

\paragraph{Compositional Multi-Axis Transfer.}
Most existing studies evaluate transfer along one dominant axis, such as cross-model, cross-domain, or cross-task transfer, even though practical source and target configurations often differ along several axes simultaneously. 
For example, an attack developed using one dataset and surrogate architecture may ultimately be applied to a different model, task, sensing modality, and deployment environment. An important open question is whether transfer along several axes can be predicted from transfer measured along the individual axes. Strong cross-model and cross-domain transfer in isolation does not necessarily imply strong transfer when both the model and domain change simultaneously.

Therefore, future work should study \emph{compositional transferability}: how attack effectiveness changes when multiple coordinates of the source configuration
$Z_s$
differ jointly from the target configuration
$Z_t$.
Systematic factorial evaluations could help identify whether particular combinations of changes amplify or suppress transfer. 
Such analysis would also clarify which shared representations, training procedures, or deployment invariances remain useful when several assumptions underlying transfer are violated at the same time.

\paragraph{Transfer Distance and Configuration Similarity.}
The current literature commonly treats transfer as a binary distinction between source and target configurations, but the magnitude of configuration difference can vary substantially. 
Two independently trained models with the same architecture may be much more similar than models belonging to different architectural families, while two datasets from related domains may differ less than datasets collected from fundamentally different environments. Similar considerations arise for tasks, modalities, and deployment conditions.

Future work needs to develop measurable notions of \emph{configuration distance} and investigate how attack effectiveness degrades as source--target distance increases. 
Such measures may incorporate representation similarity, distribution divergence, architectural differences, task relationships, preprocessing differences, or physical-channel variation. 
A central theoretical question is whether transferability decreases smoothly with increasing configuration distance or instead exhibits threshold effects in which small changes to particular components sharply reduce attack success. Better characterization of this relationship could also guide surrogate selection by identifying which source configurations provide the most informative approximations to unknown targets.

\paragraph{Temporal and Model-Version Transfer.}
Transferability should also be studied across time. Production AI systems are continuously modified through model updates, fine-tuning, safety alignment, retrieval changes, preprocessing revisions, and deployment-specific optimization. Consequently, an attack that transfers to a target at one point in time may fail after a model or system update even when the nominal task and interface remain unchanged.

Future studies are expected to evaluate whether transferable artifacts persist across model checkpoints, fine-tuning stages, software versions, and safety updates. For foundation models and LLM-based systems, this includes transfer across changes to alignment procedures, system prompts, moderation layers, and other guardrails. For perception systems, preprocessing pipelines, sensor calibration, and inference implementations may similarly change over time. Longitudinal evaluation would clarify whether transferable vulnerabilities persist across generations of related systems or whether routine updates substantially alter the relevant attack surface.

\paragraph{Standardized Adaptation and Query Accounting.}
Another unresolved issue is how to distinguish transferability from target-specific optimization. A fixed artifact applied unchanged to an unseen target represents a stronger form of transfer than an attack that uses target outputs to rank candidates, recalibrate parameters, or repeatedly refine the artifact. Both settings are practically relevant, but they should not be reported as equivalent evidence of transferable attacks.

Future work should adopt standardized reporting of target information and adaptation budgets. At minimum, studies should state the number and type of victim queries, whether target outputs are labels, probabilities, generated responses, or internal representations, which stages of attack construction use target feedback, and whether the final artifact is modified for each target. Evaluation should distinguish zero-query transfer, query-assisted transfer, and target-optimized attacks. For training-time attacks, analogous reporting should describe control over training samples, pretrained components, model parameters, and victim adaptation procedures.

\paragraph{Uncertainty and Statistical Evaluation Across Targets.}
Transferability is typically evaluated on a small collection of target models or datasets, and reported averages can conceal substantial variation among targets. This limitation is particularly important when attack success depends strongly on architecture, initialization, alignment procedure, judge model, physical condition, or other target-specific factors. High transfer success on a few selected targets does not necessarily imply that an attack will generalize to a broader target population.

Future evaluations need to treat target models, datasets, tasks, or deployments as a population over which uncertainty must be quantified. Studies should report variability across targets and repeated runs in addition to mean attack success, together with confidence intervals or other appropriate uncertainty measures. For attacks evaluated with learned judges, such as LLM jailbreaks, sensitivity to the evaluator should also be reported. Comparative studies should separate attack-induced changes from the unattacked target baseline and should avoid selecting only the best-performing trials or targets. These practices would provide stronger evidence for claims about general transferability rather than success on particular source--target pairs.

\paragraph{Physical Transfer and Reproducibility.}
Cross-deployment transfer remains particularly difficult to evaluate because physical channels introduce variation beyond the learned model. Lighting, viewpoint, occlusion, background noise, propagation conditions, sensor characteristics, preprocessing, and hardware can all modify the attack artifact before it reaches the model. Transformation-based optimization can improve robustness to selected physical variations~\cite{athalye2018synthesizing}, but robustness to transformations represented during optimization does not establish transfer to unseen physical conditions.

Future physical-attack studies are expected to clearly document the delivery channel and separate different levels of physical evidence. Simulated room responses, for example, evaluate selected acoustic transformations, whereas over-the-air recordings additionally test playback and capture. Cross-device experiments further evaluate hardware variation. Visual attacks should similarly distinguish simulated transformations from recaptured physical artifacts and from experiments that vary cameras, distances, viewpoints, and illumination. Studies of ultrasound-based attacks~\cite{zhang2017dolphinattack, yan2020surfingattack} and visual camouflage approaches such as TPatch~\cite{zhutpatch} also illustrate that perceptibility and detectability should be evaluated alongside attack success.

\paragraph{Benchmarking Transferability.}
A standardized benchmark should reflect the multi-axis nature of transfer rather than evaluating only a fixed collection of source and target models. Existing evaluations frequently concentrate on cross-model transfer within a single dataset and task, while practical systems may simultaneously involve domain, modality, task, and deployment changes. Beyond images, text, audio, graphs, and video, sensing systems also use signals such as accelerometers, gyroscopes, and LiDAR~\cite{wang2018socialite, wang2023patch, li2021deep}.

Future benchmark suites should specify source--target configuration pairs across the six transfer axes, include heterogeneous model families and datasets, and explicitly control attacker knowledge and adaptation budgets. Benchmarks should report both single-axis and multi-axis transfer so that interactions between configuration changes can be studied. 
Benchmarks should also include successful and failed transfer cases rather than selecting only settings in which attacks generalize well. Common reporting of attack success, transfer gaps, benign utility, target-query budgets, computational cost, and physical constraints would enable more meaningful comparison across methods. Such benchmarks could provide the empirical foundation needed to test theoretical explanations of transferability.

\paragraph{Transfer-Aware Defense.}
Finally, defenses should be evaluated against transferable vulnerabilities rather than only against attacks optimized for a single defended model. A defense may suppress an attack constructed directly against one model while leaving vulnerabilities that remain shared across related architectures, datasets, or deployments. Conversely, preprocessing or model modifications may reduce one form of transfer while unintentionally creating new shared failure modes.

Future defenses should be evaluated on held-out attacks and held-out target configurations, with unchanged transferred artifacts distinguished from attacks subsequently adapted to the defense. Defense evaluation should also consider multiple stages of the AI pipeline, including sensing, preprocessing, learned representations, model inference, retraining, and downstream decision-making. Combining robust model design, sensor-level protection, anomaly detection, redundancy, fail-safe mechanisms, and monitoring may provide defense in depth, but the contribution of each layer should be evaluated against explicit transfer axes.

\section{Conclusion}\label{conclusion}
This survey organizes transferable attacks across learning-based attack families through a unified six-axis taxonomy: cross-instance, cross-domain, cross-modality, cross-model, cross-task, and cross-deployment. 
By connecting transfer mechanisms and enhancement methods with conditions for success and failure, this work provides a common foundation for investigating how vulnerabilities extend beyond individual systems. 
The identified priorities in reproducible evaluation, benchmarking, adaptation accounting, and transfer-aware defense can guide future research and practical security assessment. 
More broadly, this work can help researchers and practitioners identify shared failure modes, assess risks as models and deployments evolve, and develop defenses that protect privacy, preserve system integrity, and improve safety in AI-enabled applications.


% \begin{acks}
% We would like to thank the anonymous reviewers for their insightful comments on our work. This work was supported in part through National Science Foundation grants CNS-1950171. 
% \end{acks}

\bibliographystyle{ACM-Reference-Format}
\bibliography{references}

% \appendix
% \section{Appendix}
% \input{chapters/appendix}

\end{document}